\documentclass{article}

\usepackage[utf8]{inputenc}
\usepackage[hidelinks]{hyperref}
\usepackage[numbers]{natbib}
\usepackage{amsmath}
\usepackage{xcolor}
\usepackage{graphicx}
\usepackage{booktabs}
\usepackage{algorithm}
\usepackage{algpseudocode}
\usepackage{multirow}

\algrenewcommand\algorithmicrequire{\textbf{Input:}}
\algrenewcommand\algorithmicensure{\textbf{Output:}}

\definecolor{red}{rgb}{1,0,0}
\begin{document}
    \title{Probabilistic modeling of delays for train journeys with transfers}
    \author{Nikolaus Stratil-Sauer$^a$ (nikolaus.stratil-sauer@liu.se) 
        \and Nils Breyer$^a$\footnote{Corresponding author.} (nils.breyer@liu.se) 
    }
    \date{    $^a$Department of Science and Technology, Linköping University \\[2ex]%
        \today
    }
    \maketitle
    
    \begin{abstract}
    Reliability plays a key role in the experience of a rail traveler. The reliability of journeys involving transfers is affected by the reliability of the transfers and the consequences of missing a transfer, as well as the possible delay of the train used to reach the destination. In this paper, we propose a flexible method to model the reliability of train journeys with any number of transfers. The method combines a transfer reliability model based on gradient boosting responsible for predicting the reliability of transfers between trains and a delay model based on probabilistic Bayesian regression, which is used to model train arrival delays. The models are trained on delay data from four Swedish train stations and evaluated on delay data from another two stations, in order to evaluate the generalization performance of the models. We show that the probabilistic delay model, which models train delays following a mixture distribution with two lognormal components, allows to much more realistically model the distribution of actual train delays compared to a standard lognormal model. Finally, we show how these models can be used together to sample the arrival delay at the final destination of the entire journey. The results indicate that the method accurately predicts the reliability for nine out of ten tested journeys. The method could be used to improve journey planners by providing reliability information to travelers. Further applications include timetable planning and transport modeling.
    \end{abstract}

\newpage

\section{Introduction}
\label{sec:intro}

Reliability plays a key role in the experience of a rail traveler. The willingness of travelers to travel by train decreases when reliability is expected to be low \citep{VanLoon:reliabilitydemand,rietveld:reliabilityvaluation}. For journeys involving more than one train, travelers face additional uncertainty due to the risk of missing a transfer and the consequences for the onward journey \citep{yap:interchanges,chowdhury:bytesmotstnd}. Despite the importance of reliability for travelers, journey planning services usually return itineraries according to the schedule without any information on the reliability to be expected for a particular journey. A better understanding of the reliability of whole journeys is also a key to inform planners and railway operators to be able to make decisions improving the reliability perceived by travelers.

In this paper, we characterize the reliability of a journey using the distribution of the arrival delay at the final destination. The delay distribution gives detailed information about the deviations from the scheduled arrival at the final destination that are expected for a given journey. Three main factors impact arrival delay: 

\begin{enumerate}
    \item the probability that a transfer between trains during the journey is reached,
    \item the consequence in case of missing a transfer, and
    \item the arrival delay of the last actual train used to reach the destination.
\end{enumerate}

In this paper, we aim to predict the arrival delay distribution for a given train journey with transfers. This distribution indicates the reliability of a journey by showing the probability and magnitude of deviations compared to the scheduled arrival that a traveler may expect. We use historic railway traffic data from Sweden to train and validate the method.

The main contribution of this paper is a flexible method to estimate the reliability of train journeys, including journeys with one or more transfers. As part of this method, we propose a transfer model to estimate the reliability of a transfer between trains using gradient boosting, taking into account whether a previous transfer has been missed. We propose a probabilistic model of train arrival delays using a mixture lognormal distribution that outperforms a standard lognormal model. The proposed method combines both models to predict the reliability of train journeys. We validate the predictions using historic train traffic data from several train stations in Sweden. The method could be used to highlight unreliable journeys and inform both travelers and decision makers, for example by incorporating them in journey planner applications.

The remainder of the paper is organized as follows. In Section \ref{sec:previous-literature}, we review related work. Section \ref{sec:data} describes the dataset, including its sources, structure, and pre-processing steps. In Section \ref{sec:method}, we outline the proposed method, detailing the models we developed. Section \ref{sec:results} presents the results, highlighting key findings and their implications. In Section \ref{sec:discussion}, we discuss the results, their implications and address the limitations of the method. Finally, Section \ref{sec:conclusions} concludes the paper by summarizing the main contributions and suggesting directions for future research.

\section{Related Work}
\label{sec:previous-literature}

Extensive research has been conducted on the prediction and modeling of train delays \citep{tiong:delays}. These approaches can generally be divided into long-term versus short-term predictions and event-driven versus data-driven predictions \citep{Spanninger2022}. Short-term predictions involve real-time data and are used to inform travelers currently traveling. Real-time data is not available in the case of long-term predictions, which can be used to provide information during the route planning and ticket booking stages. Event-driven approaches incorporate a dependency structure between train events (arrivals and departures) and construct a network of consecutive events. Graph-based models such as Bayesian networks \citep{corman:delays} have been used to model this problem among others. Data-driven approaches, on the other hand, have shown increasing interest in recent years \citep{tiong:delays}. These methods predict the train delay directly, without modeling a dependency structure or intermediate predictions. Probability distributions like the exponential, lognormal or Weibull distributions can be used to model train delays \citep{markovic2015analyzing}. Data-driven approaches often use machine learning methods such as neural networks \citep{luo2023multi} or decision tree-based methods \citep{shi:prediction}.

While the previously mentioned approaches show good performance for predicting train delays at a certain station or for a single leg, they do not consider complete journeys with transfers between trains. \citet{bates2001valuation} reviewed the concepts of reliability and travel time variability for public transport journeys with transfers from a theoretical perspective. Few studies have modeled the reliability of journeys with transfers using real-world data. \citet{keyhani:arriveintime} proposed a method that computes the reliability of train journeys with transfers using a probabilistic graph model based on empiric observations of train delays. \citet{stratilsauer:arrivaldelay} proposed a framework for modeling the arrival delay of train journeys with one transfer using a probabilistic Bayesian regression model. The problem of finding a reliable and robust route for a given journey has been studied by \citet{bohmova:historical} and \citet{muller:connections} among others. \citet{liu:connectivtyreliability} investigated the connectivity reliability of urban transport networks, identifying important interchange stations and causes of delays in the network. 

The reliability of a journey can be described in different ways. \citet{bates2001valuation} uses the arrival time distribution to illustrate the probability of different delays at the destination. While this is a comprehensive way to describe the reliability of a single journey, it is difficult to communicate to travelers and hard to compare between journeys. A simpler reliability metric called \textit{reliability rating} was introduced by \citet{keyhani:reliabilityconnections}. It specifies in percent how many cases a traveler can expect to reach all transfers on a journey and arrive at the destination with the train they intended to take. \citet{dixit:passengerreliablity} extends the \textit{reliability buffer time} metric first introduced by \citet{chan:od} and \citet{uniman:reliability} to journeys with multiple legs. This metric measures the variability in travel time a traveler can expect. It is defined as the difference between an extreme $n$th percentile (usually set to 95) and the 50th percentile travel time and can be interpreted as the additional time passengers have to account for during their journey to ensure arrival during this time period 19 out of 20 times.

\section{Data}
\label{sec:data}

\begin{figure}[htb!]
    \centering
    \includegraphics[width=0.6\linewidth]{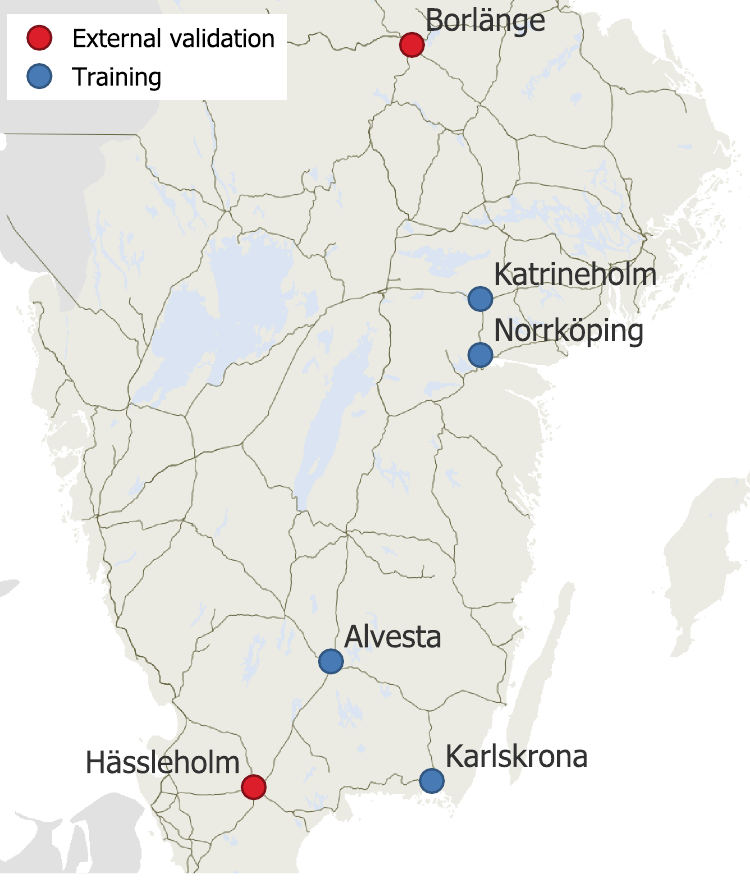}
    \caption{Location of the stations used in model training and validation with the Swedish railway network according to jnbkarta.trafikverket.se in the background}
    \label{fig:sweden_data_stations}
\end{figure}

The dataset used in this paper consists of train arrival and departure information of passenger trains from multiple train stations in Sweden. The data were collected using an API provided by the Swedish Transport Agency Trafikverket \footnote{\url{https://data.trafikverket.se/}} between December 2022 and August 2024. The stations of Alvesta, Norrköping, Karlskrona and Katrineholm are used to train the models. Observations from the stations of Hässleholm and Borlänge form the external test dataset used to judge the model performance on stations on which it was not trained. This is important to validate the generalization performance of the model \citep{tiong:delays}. Figure \ref{fig:sweden_data_stations} shows the location of the selected stations in the Swedish rail network. Norrköping, Alvesta and Hässleholm are important hubs for trains on Sweden's southern mainline connecting Malmö and Stockholm, and many smaller connecting regional lines. Katrineholm is included because it lies on the western mainline that connects Stockholm and Göteborg. Karlskrona and Borlänge are included in the training and test data sets, respectively, because they are located off the main train lines and thus provide train events with different characteristics than the other stations.

Each observation in the dataset denotes a train stopping at a station and contains the following information:

\begin{itemize}
    \item Train metadata (operator, origin, destination, train type)
    \item Scheduled time of arrival and departure
    \item Actual time of arrival and departure
\end{itemize}

Trains are separated into two types: regional trains and intercity (long-distance) trains. Since we focus on journeys that can be completed on a single day, night trains are excluded, as well as special trains such as museum trains. Intercity trains are all trains denoted \textit{IC} or \textit{Snabbtåg} by the state-owned operator SJ and all trains by the open-access operators Snälltåget and VR. All other trains are regional trains.

The observations only contain the events for a specific station. To obtain additional information about the train's route, these observations are joined with timetable data following the General Transit Feed Specification (GTFS) \footnote{\url{https://www.trafiklab.se/api/gtfs-datasets/gtfs-sweden/}}. We use these data to obtain each train's \textit{runtime} from its origin to the current station and also the \textit{total runtime} between the origin and the final destination. 

In a few cases, this join is unsuccessful and no or faulty runtime information was obtained therefore these observations are discarded. Observations with missing or faulty delay information are filtered out as well. In total, this filtering process removes 9\% of the observations. 159\,266 observations remain that are distributed as follows between the stations: 59\,356 in Alvesta, 6\,159 in Karlskrona, 33\,056 in Katrineholm and 60\,095 in Norrköping. Of these observations we randomly select 80\% as training data (127\,302 observations) and the remaining 20\% as the regular test dataset  (31\,964 observations). The external test dataset consists of 6\,000 observations in Hässlehom and 6\,000 in Borlänge. The models are evaluated on both the regular and the external test datasets.

\subsection{Transfer data}

Each transfer consists of an arrival event for the incoming train and a departure event for a train to which the traveler may transfer. All trains scheduled to leave the station between 3 and 60 minutes after an incoming train is scheduled to arrive are considered a transfer. This leaves us with 692\,191 transfers in the training dataset and 173\,421 in the test dataset derived from the train arrival and departure data. The external test dataset consists of 13\,309 observations.

The \textit{PlannedTransferTime} (PTT) for a transfer is defined as the time between the arrival of the incoming train and the departure of the connecting train, according to the schedule. A transfer is considered as \textit{Reached} if there are at least 3 minutes between the actual arrival of the incoming train and the actual departure of the connecting train. This limit is set to account for the time it takes to alight the train and walk to another platform.

\subsection{Exploratory Data Analysis}
\label{sec:EDA}

\begin{figure}[htb]
    \centering
    \includegraphics[width=0.8\linewidth]{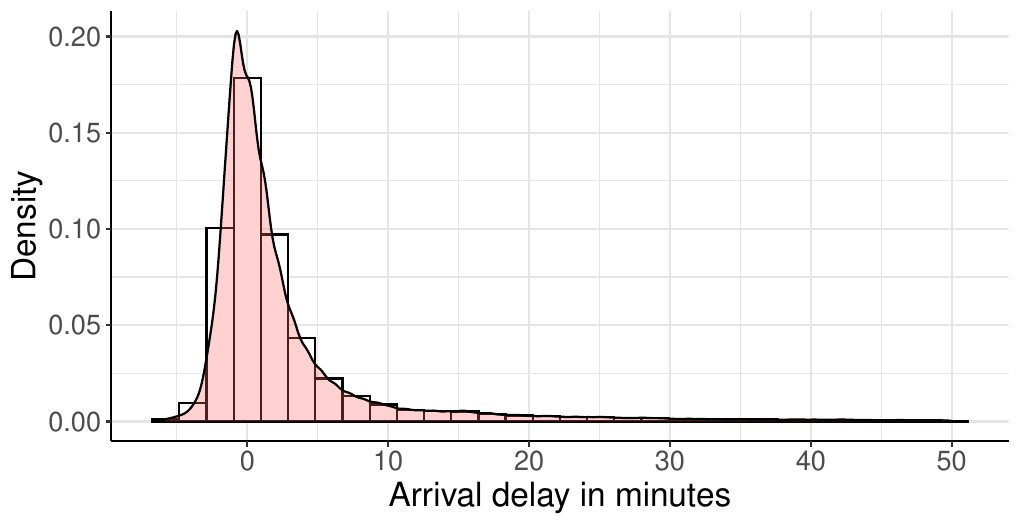}
    \caption{Distribution of arrival delays in minutes. Total sample size of 159\,266 observations, including 2\,787 observations with delay higher than 50 minutes that were removed for better readability.}
    \label{fig:delay_distribution}
\end{figure}

Train delays of passenger trains generally show a right-skewed distribution. Although some trains arrive slightly early, many more trains arrive delayed and the delays can occasionally be very large. Figure \ref{fig:delay_distribution} and Table \ref{tab:delay_statistics} show that the maximum delay in our dataset is more than 7 hours, while the mean delay is 4 minutes. The right- skewed nature of the distribution is confirmed by the fact that the mean is higher than the median and the positive skewness of 6.996.

\begin{table}[htb]
    \centering
    \begin{tabular}{lcccccc}
        \toprule
        \textbf{Min} & \textbf{Median} & \textbf{Mean} & \textbf{Max} & \textbf{Var.} & \textbf{Skewness} & \textbf{Kurtosis} \\ 
        \midrule
        -5.98333     & 0.5333          & 4.117         & 376.267      & 188.209       & 6.996             & 79.2104           \\ 
        \bottomrule
    \end{tabular}
    \caption{Summary statistics of the arrival delay (in minutes).}
    \label{tab:delay_statistics}
\end{table}

Key patterns and interactions between the target variables (arrival delay and transfer reliability) and the available features are outlined in the following paragraphs.

\begin{figure}[htb]
    \centering
    \includegraphics[width=\linewidth]{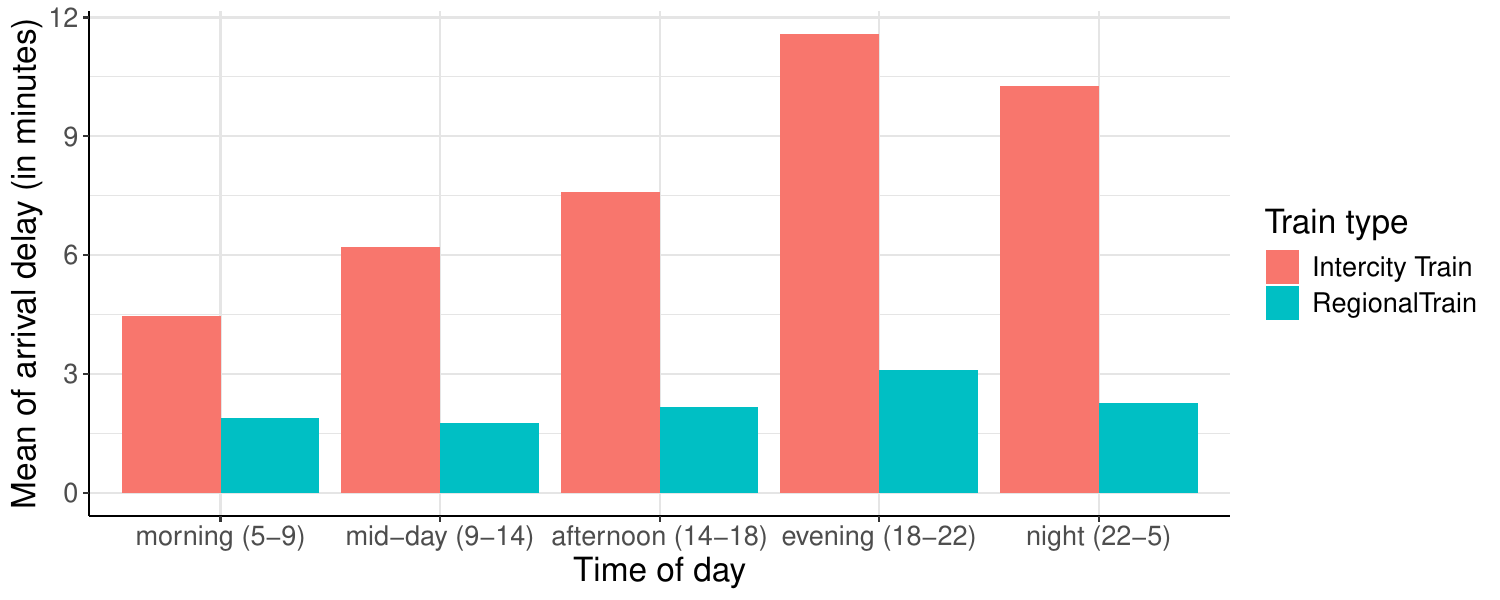}
    \caption{Train arrival delay per train type and time of day.}
    \label{fig:delay_time_type_interaction}
\end{figure}

Figure \ref{fig:delay_time_type_interaction} shows that there is a correlation between the features time of the day and train type and the arrival delay. Intercity trains have higher delays in general and the average delay for these trains increases throughout the day, while the impact of the time of the day on the delay of regional trains is much smaller. This indicates an interaction effect between these features.

\begin{figure}[htb]
    \centering
    \includegraphics[width=0.8\linewidth]{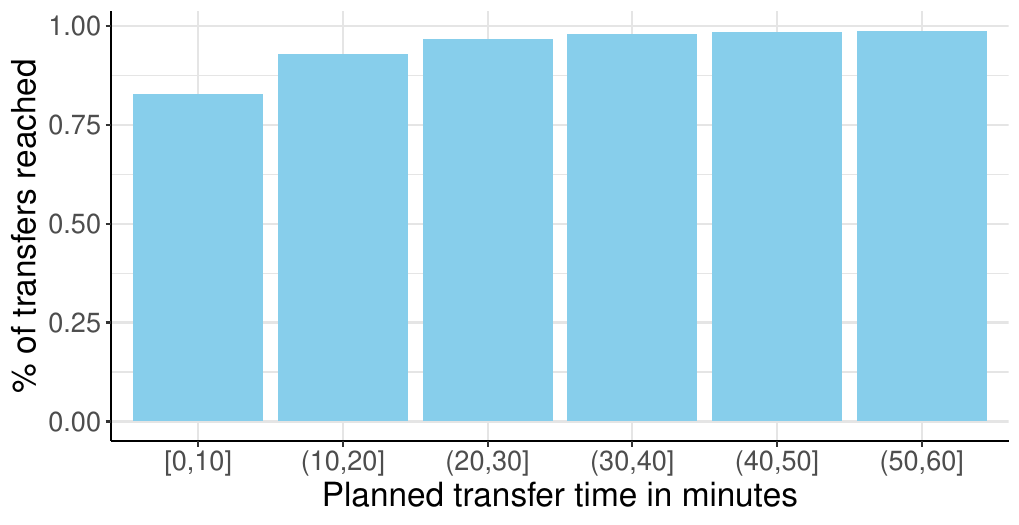}
    \caption{Transfer reliability depending on the PTT.}
    \label{fig:connection_reliability_ptt}
\end{figure}

Figure \ref{fig:connection_reliability_ptt} shows that the reliability of transfers increases drastically when the available transfer time is longer. A short transfer time of less than 10 minutes indicates a transfer reliability of only 80\%, transfers with more than 30 minutes of transfer show an average transfer reliability as high as 98\%. The reliability of transfers is naturally connected to the delay of a train and the predictors mentioned above are also relevant in this context. In total, 94.8\% of the transfers in the dataset are reached and only 5.2\% are missed.

\section{Method}
\label{sec:method}

In this section, we propose a method to estimate the reliability of a given train journey, including one or more transfers. The method uses two models as building blocks: A transfer reliability model that is responsible for predicting the reliability of a transfer, and a delay model which is used to model train arrival delays. Both models are trained using historical train information. Finally, we present how both models can be used together to sample the arrival delay at the final destination of the journey, describing the reliability of the journey.

Let $J$ be a train journey consisting of one or more legs. Legs are denoted as $J^i$ with $\overline{J}$ specifying the total number of legs in the journey. A journey also has a scheduled arrival time at the final station as $t_{s}(J)$. The arrival delay of a leg is denoted as $Delay(J^i)$. Two consecutive legs of a journey are connected by a transfer, which is denoted as $Tr(J^i, J^{i+1}) \in \{0,1\}$. If the transfer was missed, we set $Tr = 0$ and if it was reached $Tr = 1$. Table \ref{tab:notations} gives an overview of the notations used throughout the rest of the paper.

\begin{table}[htb]
\centering
\begin{tabular}{lp{0.6\linewidth}}
\toprule
\textbf{Notation}              & \textbf{Definition}                 \\ 
\midrule
$J $                           & A journey                           \\
$\overline{J} $                & The number of legs in the journey\\ 
$J^i $                         & The i-th leg of journey $J$, $i \in \{1,\ldots,\overline{J}\}$ \\ 
$t_{s}(J)$                 & Scheduled arrival time at the destination of $J$   \\ 
$dest(J)$                      & Destination of journey $J$   \\ 
$Tr(J^i, J^j) \in \{0,1\}$              & A transfer between leg $J^i$ and leg $J^j$, which is $1$ when the transfer is reached and $0$ if it is missed  \\ 
$Pr(Tr)$                       & Probability to reach transfer $Tr$ given no previous alternative was missed, predicted by transfer model  \\ 
$Pr(Tr|Tr_{prev})$             & Probability to reach transfer $Tr$ given that previous alternative transfer $T_{prev}$ was missed, predicted by transfer model  \\ 
$alternative(J, destination)$  & Alternative journey to $destination$ given the last leg of $J$ was missed  \\ 
$p(Delay(J^i))$                & Distribution of delays for the arrival of leg $J^i$ predicted by the delay model  \\ \bottomrule
\end{tabular}
\caption{Overview of the notations used in the description of the method.}
\label{tab:notations}
\end{table}

\subsection{Transfer model}

The \textit{transfer model} aims to predict the probability $Pr(Tr)$ of reaching a particular transfer $Tr$. It also gives the probability $Pr(Tr|Tr_{prev})$ of reaching a transfer $Tr$, conditional on an earlier, preferred transfer $Tr_{prev}$ being missed. 
 
To model the probability of reaching a transfer we use an XGBoost model \citep{chen2015xgboost}. XGBoost relies on gradient boosting of decision trees to predict the outcome of regression or classification problems. Predicting the reliability of transfers is a binary classification task, with imbalanced class proportions, since only about 5\% of the transfers in the training data are missed. Decision Tree based methods and XGBoost in particular are well suited for this problem because they can easily handle $NA$ values, work well with imbalanced data and show good performance in classification tasks when compared with other methods \citep{khan2024review}.

An XGBoost model has several hyperparameters including the number of trees to fit, the depth of these trees and the size of the sample used to fit each iteration, among others. The right hyperparameters depend on the given data and have to balance between optimal prediction performance and overfitting concerns. To find the optimal setup, we use grid search using cross-validation on training data.

\bigskip

\textsc{Features.} Table \ref{tab:transfer_features} presents the features used in the transfer model. It contains features about the transfer, as well as characteristics of the arriving and departing leg. Some features are combined interaction variables, motivated by the relationships found in Section \ref{sec:EDA}. To reflect the interaction between the delay, train type and hour of day, the feature \textit{arr.IntercityHour} contains this interaction inherently.

\begin{table}[htb]
\centering
\begin{tabular}{lp{0.55\linewidth}}
    \toprule
    \textbf{Feature} & \textbf{Definition} \\ 
    \midrule
    \textit{PTT} & Planned time between the arrival of the incoming and the departure of the outgoing leg \\ 
    \textit{prevPTTDiff} & Difference between the PTT of this transfer and the PTT of a previous transfer if available, NA otherwise \\ 
    \textit{Weekend} & 1 on a weekend, 0 otherwise \\ 
    \textit{arr.IntercityHour} & Interaction variable, if arriving train is an intercity train equal to the current hour (24h clock), 0 otherwise \\ 
    \textit{arr.ShortTrain} & 1 if the total runtime of the arriving train is under 2 hours, 0 otherwise \\ 
    \textit{arr.IntercityWinter} & Interaction variable, 1 if it is winter (December - February) and the arriving train is an intercity train, 0 otherwise \\ 
    \textit{dep.IntercityTrain} & 1 if the departing train is an intercity train, 0 otherwise \\ 
    \bottomrule
\end{tabular}
\caption{Features used in the transfer model.}
\label{tab:transfer_features}
\end{table}

The feature $arr.ShortTrain$ is used instead of including the runtime directly as a feature, because this led to overfitting issues, where the model would interpret it as categorical instead of continuous, resulting in unrealistic predictions. For example, for a sample with a runtime of the incoming train of 5 hours, the model predicted a 90\% transfer reliability. But just changing the runtime to 4.9 hours, while leaving all other features the same, resulted in a prediction of 71\% transfer reliability. This behavior is most likely due to clusters in the training data at certain runtime values and not due to any real and explainable patterns. 

Due to similar issues with the PTT, a monotone constraint was enforced on this feature in the XGBoost model definition. This step ensures that a higher PTT always results in a equal or higher predicted transfer reliability.

\bigskip

\textsc{Model evaluation.} Imbalanced classification requires special consideration when selecting the evaluation metrics. Accuracy is not a suitable metric, since even a baseline model, that always predicts transfers to be reached would achieve a close to perfect 95\% accuracy. AUROC (Area Under the Receiver Operator Characteristic) plots the true positive rate against the false positive rate and scores the area under the resulting curve \citep{ling2003auc}. A baseline model, choosing randomly between the classes has an AUROC of 0.5. A perfect model has an AUROC score of 1. Given a random positive and negative sample, the AUROC score can be interpreted as the rate of correctly predicting a higher probability for the positive sample. While the AUROC metric can be used on imbalanced classification problems, it does not directly evaluate the goodness of the predicted probability labels \citep{lobo2008auc}. To ensure proper calibration of these labels and the real observed probabilities we use a calibration curve visualization, which compares the predicted probability labels with the actual observed probabilities. If the predicted probabilities are calibrated well, they follow the diagonal line closely. This visualization is also known as a reliability diagram \citep{niculescu2005predicting}. Calibration curves and AUROC scores are computed for both the regular test dataset and the external test dataset, to evaluate the generalization performance of the model.

\subsection{Delay model}

The \textit{delay model} gives a probabilistic estimate of the arrival delay \\$Delay(J^i)$ in the form of a sample of delays following a probability distribution.  \citet{markovic2015analyzing} notes the difficulty of fitting a single distribution, such as a lognormal or exponential distribution to train delays. We could confirm this for our data (see Section \ref{sec:results-delay-model}) and found that a single distribution could not capture trains with very high delays properly. Therefore, we propose using a probabilistic Bayesian regression model, based on a mixture distribution with two lognormal components, which is better suited to model train delays \citep{stratilsauer:arrivaldelay}. 

A Bayesian model approximates the underlying distribution of the data by treating parameters such as the mean or the variance as probability distributions rather than fixed values. Instead of estimating a single best-fit parameter, it samples from the posterior distribution of these parameters, allowing for a more flexible representation of the variability in train delays. In a Bayesian regression model, the features can be used to model any number of distribution parameters instead of sampling them directly. This is done by including regression coefficients for each feature and sampling a distribution for each coefficient \citep{gelman2013bayesian}.

In our model for train delays, the features are used to model the mean and the variance of both lognormal components of the mixture model:

\[
p(x) = \pi_1 \cdot p_1(x \mid \mu_1, \sigma_1) + \pi_2 \cdot p_2(x \mid \mu_2, \sigma_2),
\]
where:
\[
p_i(x \mid \mu_i, \sigma_i) \sim \text{lognormal}(x \mid \mu_i, \sigma_i),
\]
\[
\mu_1 = \beta_{1,0} + \beta_{1,1} X_1 + \beta_{1,2} X_2 + \cdots + \beta_{1,k} X_k,
\]
\[
\mu_2 = \beta_{2,0} + \beta_{2,1} X_1 + \beta_{2,2} X_2 + \cdots + \beta_{2,k} X_k,
\]
\[
\sigma_1 = \exp(\beta_{3,0} + \beta_{3,1} X_1 + \beta_{3,2} X_2 + \cdots + \beta_{3,k} X_k),
\]
\[
\sigma_2 = \exp(\beta_{4,0} + \beta_{4,1} X_1 + \beta_{4,2} X_2 + \cdots + \beta_{4,k} X_k).
\]

Using Markov Chain Monte Carlo (MCMC) sampling and the R package brms \citep{burkner:brms}, coefficients $\beta$ and mixing weights $\pi$ are sampled. As this is a Bayesian model, we need to define priors. All priors are set to weakly informative $N(0,1)$ priors, which have a regularizing effect, resembling ridge regression in a frequentist context \citep{lemoine2019moving}.

The main benefit of the Bayesian regression model for this problem is its probabilistic nature. Its output takes the form of samples following a probability distribution giving detailed characteristics of the uncertainty of the delay. Many other machine learning algorithms would produce point estimates of the predicted delay only. This benefit is especially relevant in the present context of long-term predictions, where the uncertainty is still high, making point estimates less meaningful.

\bigskip

\textsc{Features.} Features used in the model have to be encoded numerically and since the model is based on linear regression, all non-linear terms and interactions have to be modeled explicitly. The final accepted model uses the features shown in Table  \ref{tab:delay_features}.

\begin{table}[htb]
\centering
\begin{tabular}{lp{0.55\linewidth}}
    \toprule
    \textbf{Feature} & \textbf{Definition} \\ 
    \midrule
    \textit{TotalRuntime} & Runtime of the train from its origin to its destination (in hours) \\ 
    \textit{MidDayAfternoonIntercity} & Interaction variable, 1 if the time is between 9:00 and 18:00 and the arriving train is an intercity train, 0 otherwise \\ 
    \textit{EveningNightIntercity} & Interaction variable, 1 if the time is between 18:00 and 05:00 and the arriving train is an intercity train, 0 otherwise \\ 
    \bottomrule
\end{tabular}
\caption{Features used in the delay model.}
\label{tab:delay_features}
\end{table}

\bigskip

\textsc{Model evaluation.} In Bayesian regression models using MCMC sampling, it is important to ensure convergence of the final model. The metrics $\hat{R}$ and effective sample size $ESS$ are used to guarantee this. Depending on the problem, Bayesian models can struggle to achieve convergence, especially when presented with many features and a big sample size. If these metrics show convergence of the model, it can be accepted \citep{gelman2020bayesian}. Accepted models can be compared according to their expected log-pointwise predictive density (ELPD) on the test dataset \citep{Vehtari2017}. In addition to this, posterior predictive checks can be used to judge the calibration of the model to actual data. If the model is a good fit, it should be able to generate data that cannot easily be distinguished from actual train delay data. We can generate data from the posterior predictive distribution and compare it with actual delays using QQ plots and similar tools \citep{gabry2019visualization}. These checks can also be performed on the external test data, to determine the generalization performance of the model and ensure that it can also reliably model train delays for stations that are not present in the training data. 

\subsection{Reliability of a journey}
\label{sec:method-journey}

In order to predict the reliability of a planned journey $J_{plan}$, we need to combine the transfer and delay models. Algorithm \ref{alg:sampleJourney} presents the structure of how both models are combined to produce a delay sample of a journey, including potential transfers. At first, we iterate through the transfers in order to sample an actual journey $J_{act}$. For each transfer, the transfer model is used to predict the probability of reaching it. This probability is then used to sample if the transfer was reached or not. If it was not reached, we replace the current journey $J$ with an alternative journey starting at the station where the transfer was missed.

\begin{algorithm}[htb] 
\caption{Sample a delay $DelaySample$ given a planned journey $J_{plan}$}
\label{alg:sampleJourney}
\begin{algorithmic}[1]
\Require Journey $J$
\Ensure A delay sample $DelaySample$

\State $J \gets J_{plan}$
\State $J_{act} \gets [J_{plan}^1]$
\State $Tr_{prev} \gets NA$

\While{$dest(J_{act}^{\overline{J}_{act}}) \neq dest(J^{\overline{J}})$}
\Comment{Check if we are at final destination}

    \State $Tr_{next} \gets Tr(J^{\overline{J_{act}}},J^{\overline{J_{act}}+1})$

    \If{$Tr_{prev} = NA$}
        \State $P_{next} \gets Pr(Tr_{next})$ \Comment{using the transfer model}
    \Else
        \State $P_{next} \gets Pr(Tr_{next}|Tr_{prev}=0)$ \Comment{using the transfer model}
    \EndIf
    
    \State Draw $r \sim \text{Bernoulli}(P_{next})$

    \If{$r=1$} \Comment{Transfer reached, add to journey}
        \State $J_{act} \gets J_{act} \cap [J^{\overline{J_{act}}+1}]$
        \State $Tr_{prev} \gets NA$  
    \Else \Comment{Transfer missed, update planned journey with alternative}
        \State $Tr_{prev} \gets Tr_{next}$  
        \State $J \gets alternative(J_{act} \cap [J^{\overline{J_{act}}+1}],dest(J))$  
    \EndIf
\EndWhile

\State Draw $FinalDelay \sim p(Delay(J_{act}^{\overline{J}_{act}}))$ \Comment{using the delay model}
\State $DelaySample \gets t_{s}(J_{act}) - t_{s}(J_{plan}) + FinalDelay$ \\
\Return{$DelaySample$}
\end{algorithmic}
\end{algorithm}

\begin{figure}[htb!]
    \centering
    \includegraphics[width=\linewidth]{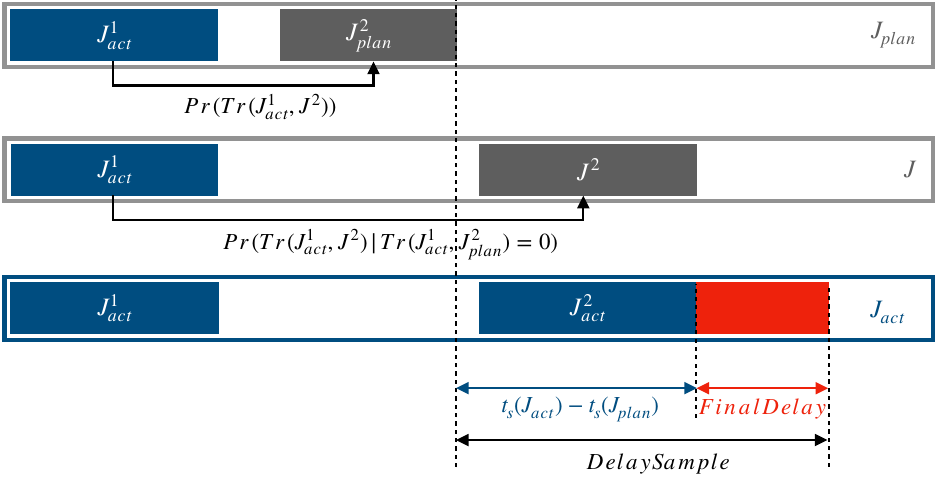}
    \caption{Example of the sample generating process for a planned journey $J_{plan}$ with one transfer. In the first two steps the transfer model is used to sample the actual journey $J_{act}$ shown in blue. The planned transfer is missed and an alternative train for the second leg has to be taken. Finally, the delay model is used to sample a final delay for the last train used shown in red. Together, the output of both models yields one $DelaySample$.}
    \label{fig:example_journey}
\end{figure}

The probability of reaching a transfer is handled conditionally if a previous transfer at the same station was missed through the feature $prevPTTDiff$ in the transfer model. It is set to the difference between the transfer time of the previous transfer and the current transfer time. It is set to $NA$, if there is no information on previous transfers. This feature represents the fact that a traveler would only consider an alternative transfer if they missed their preferred transfer. Assume that a journey has a transfer with a transfer time of 10 minutes at station $X$. In this case, $prevPTTDiff = NA$. But if this transfer is missed, there is an alternative with a transfer time of 30 minutes. Here $prevPTTDiff = 30-10  = 20$, as the knowledge about the previous missed transfer is available. This structure can be seen as a series of binary choices, as shown in Figure \ref{fig:example_journey}. It is important to include this variable because it adds relevant context to the transfer model. If a transfer with a 10-minute transfer time is missed, it is unlikely that an alternative transfer with a 12-minute transfer time would be reached. This way this information is considered and prediction performance can be improved.

The process is repeated until arrival at the final station where it results in a final delay and works with any number of transfers and alternative trains. At this point, we have sampled an actual journey $J_{act}$ which is based on both the probability of missing a transfer and the consequences for the onward journey. However, the arrival delay at the destination does not only depend on the trains reached but also on the possible delay of the train used to reach the destination. For this purpose, we use the delay model and sample an arrival delay for the final train used.

The arrival delay sample finally returned by the algorithm is the sum of $t_{s}(J_{act}) - t_{s}(J)$ and the final delay sampled by the delay model. The first term accounts for the fact that arriving with a later leg because of a missed transfer adds delay to the journey compared to the planned journey. If all transfers are reached as planned, then this term is 0. The second term accounts for the delay of the last train used. Figure \ref{fig:example_journey} illustrates this process and how a $DelaySample$ is generated for a journey with one transfer. In this example the first transfer is sampled to be missed and therefore an alternative transfer has to be used.

Details on how to obtain alternative journeys $alternative(J, destination)$ in case of a missed transfer are not in the scope of this paper. A simple approach is to use the next train connecting the same stations as the missed leg. But in principle, they could also be obtained using public transport routing APIs like the Swedish ResRobot\footnote{\url{https://www.trafiklab.se/api/trafiklab-apis/resrobot-v21/}} allowing for more flexible re-routing options. In practice, there has to be a cut-off when a journey can not reasonably be completed. This could be after e.g. 3 missed transfers in a single station. In such a case the algorithm would terminate and designate that delay as $NA$ indicating that the final station was not reached.

\bigskip

\textsc{Model evaluation.} To obtain an approximation of an arrival delay distribution for a given journey, the process is executed multiple times to obtain a representative sample of delays. This sample can be compared to actual observations of the journey using the actual arrival and departure data of the involved trains.  To evaluate the combined model structure, delays from real journeys can be compared with sampled delays. Journeys can be evaluated based on their predicted and actual reliability rating and reliability buffer time metrics (see Section \ref{sec:previous-literature}), as well as through qualitative checks of the resulting density plots.

\section{Results}
\label{sec:results}

We trained the models using the data described in Section \ref{sec:data}. In this section, we present the results including the validation of the models. At first, the transfer and delay models are evaluated one by one and finally, the output of the combined journey reliability prediction method is compared with delays from actual train journeys.

\subsection{Transfer model}

After the hyperparameter optimization, we obtain the optimal hyperparameter configuration as shown in Table \ref{tab:xgboost_params}.

\begin{table}[htb]
    \centering
    \begin{tabular}{ll}
        \toprule
        \textbf{Hyperparameters} & \textbf{Value} \\ 
        \midrule
        nrounds      & 500     \\
        max\_depth   & 5       \\
        eta          & 0.1     \\
        subsample    & 0.8     \\
        gamma        & 0.1     \\
        \bottomrule
    \end{tabular}
    \caption{Optimal XGBoost hyperparameters after CV grid search.}
    \label{tab:xgboost_params}
\end{table}

\begin{figure}[htb!]
    \centering
    \includegraphics[width=\linewidth]{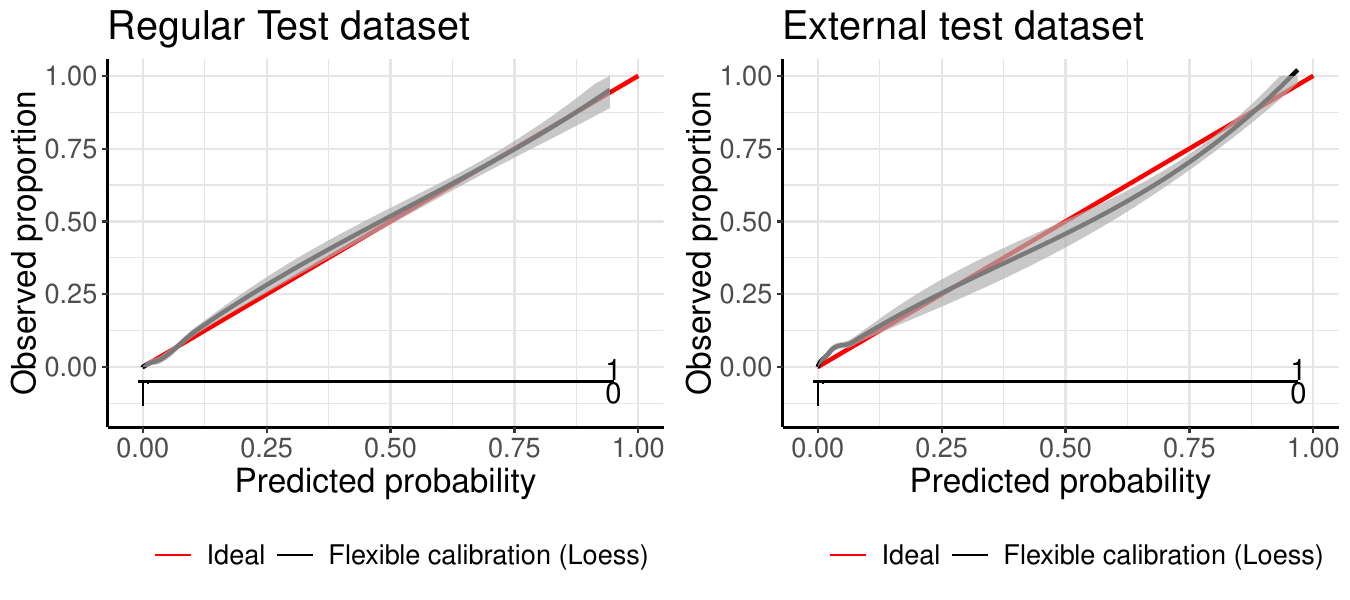}
    \caption{Comparison of calibration curves of the transfer model with the actual probabilities from the test dataset on the left and the external test dataset on the right. In this case the model predicts the probability of missing a transfer, lower values indicate higher transfer reliability.}
    \label{fig:transfer_model_cal_curve}
\end{figure}

The most important feature for the final model by the gain of predictive performance is the \\$prevPTTDiff$, followed by the $PTT$ and $arr.hourIC$ features. The AUC score on the regular test dataset is 0.943, while on the external test dataset the score is 0.905. Figure \ref{fig:transfer_model_cal_curve} shows the calibration curves of the model on the two test datasets.

The AUC scores indicate a good prediction performance. Scores are similar between the regular and external test datasets, indicating a good generalization performance on unseen data and stations that the model was not trained on. The calibration curves show that the predicted probability values resemble the true probability quite closely. In the case of the external test dataset, the model appears to be slightly too optimistic for transfers with a low probability of being missed.

\subsection{Delay Model}
\label{sec:results-delay-model}

In the Bayesian modeling workflow \citep{gelman2020bayesian} we first fit a number of models with different numbers of features, check them for convergence and then choose the preferred model based on the ELPD score. The resulting mixture model includes one component with a lower mixture weight $\pi_1$, but a higher variance $\sigma_1$ so that high delays are also given some probability mass. The other component has a higher mixture weight $\pi_2$ and lower variance $\sigma_2$ and is mainly responsible for representing trains with a lower or no delay.

\begin{table}[htb!]
\centering
\begin{tabular}{lcccc}
\toprule
\textbf{Parameter} & \textbf{Mean} & \textbf{l-95\% CI} & \textbf{u-95\% CI} \\
\midrule
\textit{$\pi_1$}                          & 0.28 & 0.28 & 0.29 \\
\textit{$\pi_2$}                          & 0.72 & 0.71 & 0.72 \\
\textit{$\mu_1$ Intercept}               & 2.03 & 2.01 & 2.05 \\
\textit{$\sigma_1$ Intercept}            & -0.45 & -0.47 & -0.42 \\
\textit{$\mu_2$ Intercept}               & 1.79 & 1.79 & 1.80 \\
\textit{$\sigma_2$ Intercept}            & -1.64 & -1.65 & -1.62 \\
\textit{$\mu_1$ TotalRuntime}           & 0.15 & 0.14 & 0.16 \\
\textit{$\mu_1$ MidDayAfternoonIntercity} & 0.03 & -0.01 & 0.07 \\
\textit{$\mu_1$ EveningNightIntercity1}  & 0.47 & 0.42 & 0.52 \\
\textit{$\sigma_1$ TotalRuntime}        & 0.04 & 0.03 & 0.05 \\
\textit{$\sigma_1$ MidDayAfternoonIntercity} & 0.35 & 0.32 & 0.37 \\
\textit{$\sigma_1$ EveningNightIntercity} & 0.26 & 0.22 & 0.29 \\
\textit{$\mu_2$ TotalRuntime}           & 0.01 & 0.00 & 0.01 \\
\textit{$\mu_2$ MidDayAfternoonIntercity} & 0.01 & 0.00 & 0.02 \\
\textit{$\mu_2$ EveningNightIntercity}  & 0.12 & 0.10 & 0.13 \\
\textit{$\sigma_2$ TotalRuntime}        & 0.13 & 0.12 & 0.13 \\
\textit{$\sigma_2$ MidDayAfternoonIntercity} & 0.19 & 0.17 & 0.21 \\
\textit{$\sigma_2$ EveningNightIntercity} & 0.35 & 0.32 & 0.38 \\
\bottomrule
\end{tabular}
\caption{Summary of the regression coefficients for the delay model. Note that parameters for $\sigma$ are on the log scale.}
\end{table}

\begin{figure}[htb!]
    \centering
    \includegraphics[width=\linewidth]{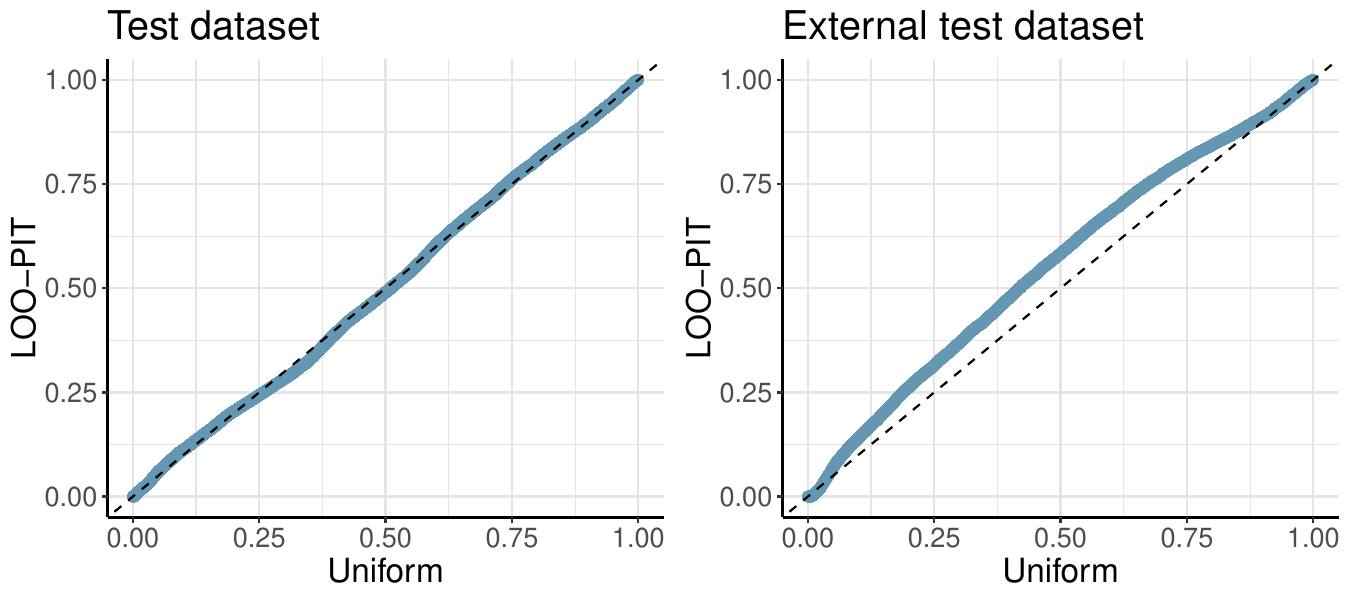}
    \caption{QQ plot comparison of the posterior predictive output of the delay model with the test and external test datasets.}
    \label{fig:delay_model_qq}
\end{figure}

To check the fit of the final model output compared with actual train delays, we can refer to posterior predictive checks in the form of QQ plots. The output of a well-calibrated model should follow the diagonal line, while deviations indicate problems. Figure \ref{fig:delay_model_qq} presents this for the regular and external test datasets. While the fit for the test dataset appears to follow the diagonal line almost perfectly, a slight deviation is apparent for the external test dataset. This concave deviation indicates that the sampled delays are slightly more left-skewed than the actual delays of the external test data.

\begin{figure}[htbp!]
    \centering
    \includegraphics[width=0.9\linewidth]{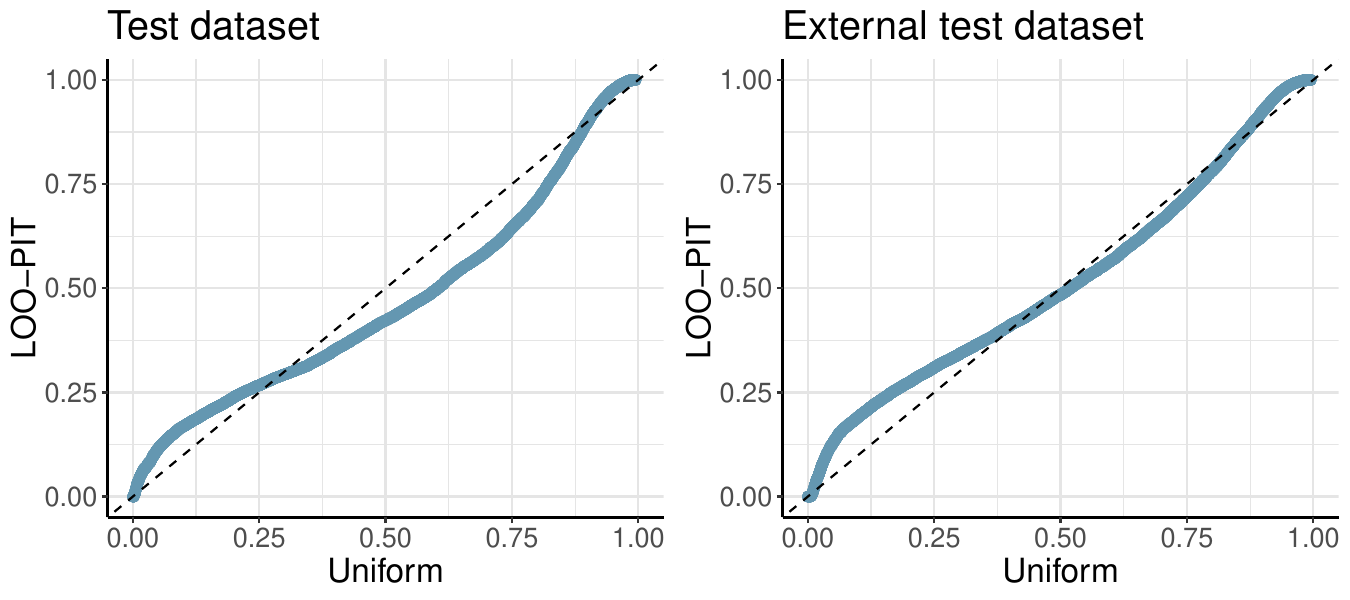}
    \caption{QQ plot of a lognormal model with just one component used on the regular test dataset.}
    \label{fig:delay_model_qq_1component}
\end{figure}

Figure \ref{fig:delay_model_qq_1component} shows the QQ plot for a model with just one lognormal component for comparison. Here we find significant deviations from the diagonal line, showing the same trend for both test datasets. This indicates that this model is not well calibrated. The S curve shows a systematic bias, where the model underestimates the uncertainty in one part of the distribution and overestimates in the other.

\subsection{Evaluating entire journeys}
\label{sec:results-entire-journey}
In order to validate the reliability predicted for whole journeys using the method presented in Section \ref{sec:method-journey}, we selected ten actual journeys (see Appendix \ref{sec:appendix-journeys}). For each day in the dataset, we get an actual delay based on the transfers reached that day and the observed delay of the last train. We compare this actual sample to a predicted sample of 1000 delays obtained by repeatedly executing Algorithm \ref{alg:sampleJourney}.

\begin{figure}[htb!]
    \centering
    \includegraphics[width=0.9\linewidth]{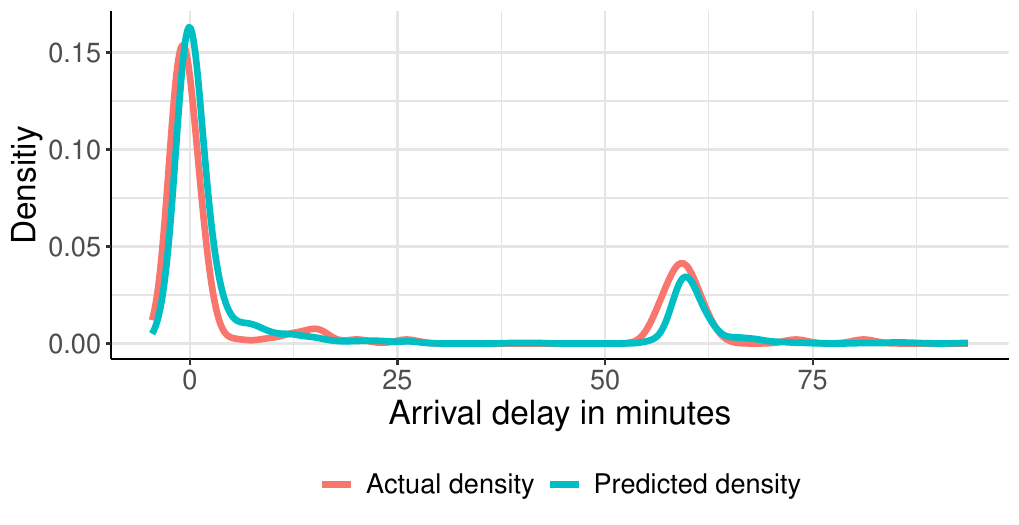}
    \caption{Comparison of the actual distribution of delays of Journey 1 with the sampled distribution of delays generated by the model.}
    \label{fig:density_comparison}
\end{figure}

Figure \ref{fig:density_comparison} shows the predicted delay distribution compared with the actual distribution of delays for Journey 1. Each peak in the distribution represents one arriving train at the final station. The first peak is the cases where all transfers have been reached and the second peak is cases where a transfer was missed and the next train, which arrives 60 minutes later, was taken.

\begin{figure}[htb!]
    \centering
    \includegraphics[width=1\linewidth]{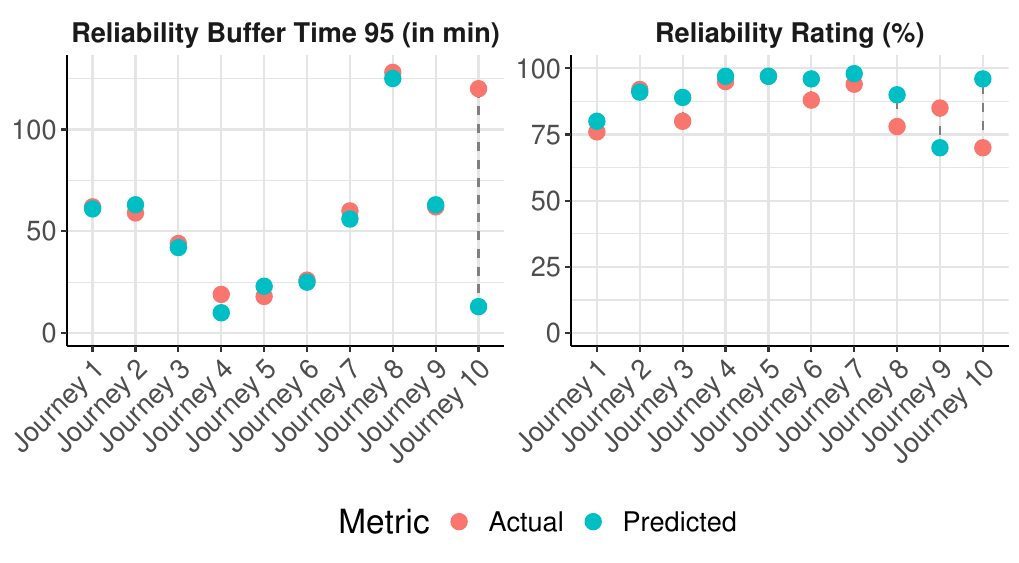}
    \caption{Comparison of the actual reliability of 10 example journeys with the predicted output of the model.}
    \label{fig:rel_rating_comparison}
\end{figure}

Figure \ref{fig:rel_rating_comparison} shows a comparison of the reliability rating and the reliability buffer time obtained from the actual sample and the sample predicted by our method for all ten journeys. The reliability rating shows how likely it is to reach the destination of the journey without missing a transfer. This metric only takes the transfer model into account, since final delays are irrelevant. The reliability buffer rating gives the difference between the 95th percentile travel time and the 50th percentile time. Here, the entire model structure is validated, since final delays as well as correctly predicting transfers is important.

In most cases, both metrics are close to the actual observed values. For journeys 8, 9, and 10 on the other hand, the error is bigger, especially considering the reliability rating. Journey 10 shows a big difference between the actual and the predicted reliability buffer time. Possible reasons behind these discrepancies for these journeys are discussed in Section \ref{sec:discussion}.

\section{Discussion}
\label{sec:discussion}

The results show that the proposed model structure can be used to model the distribution of delays and the reliability of transfers for a given train journey. While both models show good results individually, due to the makeup of the dataset and the combination of a regular classification model and a Bayesian probabilistic model, evaluating the results of the entire model structure is more difficult. We have used ten example journeys. One reason is that the existing dataset is limited to a couple of stations in Sweden. Also, often trains get a slightly modified schedule in the yearly timetable process. This makes it difficult to find stable journeys over time with enough observations.

As the final output is a distribution of delays, the selection of evaluation metrics is not straightforward. For this reason, we select evaluation metrics that are relevant to the eventual user of the model output. Both the reliability rating and the reliability buffer time carry meaningful information for a traveler and can be used to validate model outputs and compare them with actual samples.

Investigating these outcomes for the example journeys shows that the model predictions are close to the actual values in most cases, but there is a significant deviation for Journey 10. This journey includes a transfer from an intercity train to a regional train with 20 minute transfer time during the afternoon. Comparable transfers in the dataset have 90\% reliability. Yet, for Journey 10 the reliability is only 68\%. This indicates a problem that seems to be specific to this departure and therefore is hard for the model to account for.

As we can see in Figure \ref{fig:rel_rating_comparison}, in most cases the transfer model is slightly too optimistic when it comes to the reliability rating. Boosting methods like XGBoost can be prone to miscalibrated prediction outputs, where probabilities are pushed towards 0 or 1 \citep{niculescu2005obtaining}. Even though this effect is small, it is also visible in the calibration curve for the external test dataset in Figure \ref{fig:transfer_model_cal_curve}, where the model is slightly too optimistic for transfers that are likely to be reached. Implementing calibration methods like Isotonic Regression did not significantly help with this issue.


The proposed method does make some simplifications. One of these is the use of a fixed minimum transfer time, which is set to 3 minutes. This is a simplification since the exact time required to move between trains depends on the station and the exact platforms the trains arrive at, as well as the walking speed of the traveler. Some transfers involve a short walk between adjacent platforms, while others in larger stations could involve navigating sets of stairs and complex passageways. This might be improved by using a dynamic minimum transfer time depending on the tracks used by the trains and incorporating such information into the transfer model.

Network effects and interactions between trains are not fully considered by the used model structure. Delays of one train could indicate problems in the entire network, for example during a weather event. This interaction is not represented in the model structure, as both models are independent of each other. The transfer model also does not explicitly consider the possibility of two trains waiting for each other. Adding such interactions and network effects could further improve the performance, but would require a much more complex model structure.

\section{Conclusions}
\label{sec:conclusions}

We proposed a method for modeling the delays and the reliability of whole train journeys with any number of transfers. It accounts for all relevant factors influencing the arrival delay of a traveler including the possibility of missing a transfer and accounting consequences caused by this as well as the delay of the train used to reach the destination. The method is flexible and allows for example to use different ways of finding alternative routes in case of a missed transfer. 

Train delays and transfer reliability are modeled accurately overall, even though the model for transfer reliability is slightly too optimistic in some cases. The combined model also shows good performance on stations that are not part of the training dataset, indicating a capability for generalization, which is an important prerequisite for real-world usage.

The model output could be helpful for travelers during the booking and planning stage of their trip by incorporating the information into a journey planner \citep{wunderlich:visualization}. Train operators and planners could also use it to estimate the reliability of transfers and entire journeys during the timetable planning process. 

Future research is needed to extend the presented model structure for the real-time case. Real-time information on delays allows us to make much more precise statements on the expected arrival delay. This could help travelers during their trip, giving accurate information about the best alternative connection and their expected arrival time in case of a delay. Another important area to investigate is how the model could be expanded for multi-modal journeys, including other public transit modes in addition to trains. While the model structure is also applicable for these journeys, the individual models would have to be expanded, because the delay distributions would likely differ significantly between modes of transport. Reliability may affect both travel mode and route choice and an important research direction is thus to investigate how the proposed methods could be applied to improve transport models.

\bibliography{main}

\clearpage

\appendix

\section{Appendix. Detailed example journey information}
\label{sec:appendix-journeys}
\setcounter{table}{0}
\renewcommand{\thetable}{A\arabic{table}}

\begin{table}[htb!]
\centering
\begin{tabular}{@{}lllcc@{}}
\toprule
\textbf{Journey} & \textbf{Train} & \textbf{Origin} & \textbf{Destination} & \textbf{N} \\ \midrule
\multirow{3}{*}{Journey 1} & Intercity Train & Stockholm (10:21) & Alvesta (13:26) & \multirow{3}{*}{157} \\
                          & Regional Train & Alvesta (13:39) & Emmaboda (14:28) &  \\
                          & Regional Train & Emmaboda (14:34) & Karlskrona (15:16) &  \\ \midrule
\multirow{3}{*}{Journey 2} & Regional Train & Karlskrona (12:41) & Emmaboda (13:25) & \multirow{3}{*}{126} \\
                          & Regional Train & Emmaboda (13:32) & Alvesta (14:20) &  \\
                          & Intercity Train & Alvesta (14:30) & Norrköping (16:20) &  \\ \midrule
\multirow{2}{*}{Journey 3} & Intercity Train & Stockholm (16:14) & Alvesta (19:26) & \multirow{2}{*}{100} \\ 
                          & Regional Train & Alvesta (19:38) & Växjö (19:48) &  \\ \midrule
\multirow{2}{*}{Journey 4} & Regional Train & Sala (13:03) & Linköping (15:49) & \multirow{2}{*}{276} \\
                          & Regional Train & Linköping (16:04) & Mjölby (16:24) &  \\ \midrule
\multirow{2}{*}{Journey 5} & Regional Train & Älmhult (11:27) & Hässleholm (12:02) & \multirow{2}{*}{168} \\ 
                          & Regional Train & Hässleholm (12:13) & Karlskrona (14:12) &  \\ \midrule
\multirow{2}{*}{Journey 6} & Regional Train & Karlskrona (09:49) & Hässleholm (11:45) & \multirow{2}{*}{127} \\ 
                          & Regional Train & Hässleholm (11:55) & Älmhult (12:18) &  \\ \midrule
\multirow{2}{*}{Journey 7} & Regional Train & Värnamo (18:46) & Alvesta (19:18) & \multirow{2}{*}{190} \\ 
                          & Intercity Train & Alvesta (19:28) & Linköping (20:51) &  \\ \midrule
\multirow{2}{*}{Journey 8} & Intercity Train & Hässleholm (18:47) & Alvesta (19:26) & \multirow{2}{*}{66} \\
                          & Regional Train & Alvesta (19:36) & Värnamo (20:06) &  \\ \midrule
\multirow{3}{*}{Journey 9} & Intercity Train & Borlänge (9:55) & Sala (10:57) & \multirow{3}{*}{79} \\
                          & Regional Train & Sala (11:04) & Norrköping (13:23) &  \\ 
                          & Regional Train & Norrköping (13:48) & Kimstad (13:58) &  \\ \midrule
\multirow{3}{*}{Journey 10} & Regional Train & Mjölby (10:25) & Linköping (10:44) & \multirow{3}{*}{126} \\
                          & Intercity Train & Linköping (11:04) & Alvesta (12:26) &  \\ 
                          & Regional Train & Alvesta (12:46) & Rydaholm (12:57) &  \\
\bottomrule
\end{tabular}
\label{tab:detailled_journey_info}
\caption{Specification of the example journeys used for model evaluation.}
\end{table}

\end{document}